\documentclass{article}

\usepackage{microtype}
\usepackage{graphicx}
\usepackage{amsmath}
\usepackage{verbatim}
\usepackage{subfigure}
\usepackage{booktabs} 
\usepackage{hyperref}

\usepackage[accepted]{icml2021}

\icmltitlerunning{Identifying Linked Fraudulent Activities Using GraphConvolution Network}

\begin{document}
\nocite{*}
\twocolumn[
\icmltitle{Identifying Linked Fraudulent Activities Using GraphConvolution Network}



\icmlsetsymbol{equal}{*}

\begin{icmlauthorlist}
\icmlauthor{Vyom Shrivastava}{equal,to}
\icmlauthor{Sharmin Pathan}{equal,to}

\end{icmlauthorlist}

\icmlaffiliation{to}{University of Georgia, USA}

\icmlcorrespondingauthor{Sharmin Pathan}{sharmin.pathan07@gmail.com}
\icmlcorrespondingauthor{Vyom Shrivastava}{vyom.shrivastava@gmail.com}

\icmlkeywords{Account Takeover, Graph convolution networks, community detection, Gradient Boosting Trees, XGBoost, SMOTE, Label Propagation}

\vskip 0.3in
]\hyphenpenalty=10000

\begin{abstract}
In this paper, we present a novel approach to identify linked fraudulent activities or actors sharing similar attributes, using Graph Convolution Network (GCN). These linked fraudulent activities can be visualized as graphs with abstract concepts like relationships and interactions, which makes GCNs an ideal solution to identify the graph’s edges which serve as links between fraudulent nodes. Traditional approaches like community detection require strong links between fraudulent attempts like shared attributes to find communities and the supervised solutions require large amount of training data which may not be available in fraud scenarios and work best to provide binary separation between fraudulent and non fraudulent activities. Our approach overcomes the drawbacks of traditional methods as GCNs simply learn similarities between fraudulent nodes to identify clusters of similar attempts and require much smaller dataset to learn. We demonstrate our results on linked accounts with both strong and weak links to identify fraud rings with high confidence. Our results outperform label propagation community detection and supervised GBTs algorithms in terms of solution quality and computation time.
\end{abstract}

\begin{figure*}[ht]
\vskip 0.2in
  \centering
  \fbox{ \includegraphics[scale=0.4]{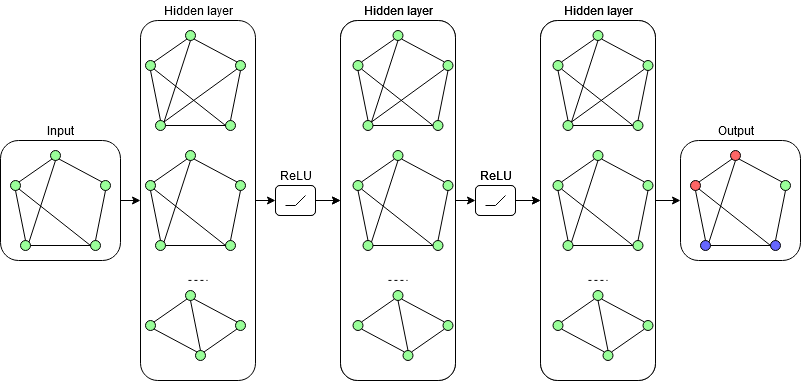} \rule[-.5cm]{0cm}{0cm} \rule[-.5cm]{0cm}{0cm}} 
  \caption{A schematic representation of our proposed network architecture. Hidden layers apply graph convolution over embeddings from previous layers. Final layer provides community/cluster labels to the nodes. \newline}
  \label{network}
  \vskip -0.2in
\end{figure*}
\section{Introduction}
\label{submission}

In 2019, the USA suffered more than \$2 Billion in losses due to fraudulent activities, estimated to increase more than 20\% in 2020 \cite{fundament}. Some of the largest frauds that contribute to this loss are identity theft, account takeovers, synthetic identities/accounts which are supported by money laundering. Fraudsters usually utilize scripts to create these accounts or takeover accounts in bulk hence sharing similar attributes, behavior or links between these attempts. Such linked attempts can be visualized as graphs with abstract concepts like relationships and interactions in the form of shared attributes and similar features therefore Graph Convolution Networks\cite{kipf2017semisupervised} are an ideal solution to identify such linked fraudulent activities.
\newline
\newline
Traditional approaches utilized solutions like unsupervised clustering algorithms such as community detection\cite{7501694} or supervised classification algorithms \cite{9103880} to identify fraudulent attempts. The main drawbacks of such solutions are that they either require strong links between attempts like sharing the same attributes such as the same cookies or device to find communities relying on the assumption that connected nodes in the graph are likely to share the same label. This assumption might miss fraud with no direct links restricting the modeling capacity, as graph edges need not necessarily encode node similarity, but could contain additional information resembling the behavioral patterns in the fraud request. On the other hand, Supervised solutions require large amount of past information of identified fraud to learn fraud patterns which may not always be available in such abundance for most systems. GCNs overcome both the drawbacks as they require very few examples of fraud to identify similar activity and learn the fraud behavior based on attributes associated with the attempt to identify similar activities even though no strong link or shared attribute may be present.
\newline
\newline
Despite requiring strong links to identify fraud, community detection \cite{7501694} solutions such as label propagation prove useful to identify rings of identical fraud. Although, a major issue with community detection solution is that propagating the rings requires a long time to run which may not be ideal with time sensitive fraud. Kipf et al. \cite{kipf2017semisupervised} proposed fast approximate spectral graph convolutions using a spectral propagation rule which results in a faster and efficient convergence and propagation. In this paper, we use the spectral graph convolution framework which provides a semi-supervised approach to identify linked fraud in form of rings which can be propagated with a few examples from each ring provided to the network to learn similarities.

\section{Problem Formulation}

In this paper we demonstrate linked fraud detection for Account Takeovers. We define a graph ${G=\{V,E\}}$ with nodes  ${V}$ representing each account that may or may not be a victim of a takeover, with edges  ${E}$ connecting two nodes, representing a link between the nodes. In our problem setting, each node is associated with a set of features ${X = \{x^i, i = 1, 2, .., n\}}$ representing login and account behavior patterns like activity timeout, mouse movement, login velocity, distinct cookie count, keystroke speed, account tenure, time since last active, IP velocity, IP risk score etc. 
\newline
\newline
In our problem setting, total fraud is less than 5\% with known fraud less than 15\% of the total fraud in the system. An activity is considered Account TakeOver (ATO) if the attempt is not from the account owner but from an unidentified actor not related to the account owner. The known or identified fraud is usually from users notifying about unauthorized access to their account with some notable changes unrecognized by the account owner. A node is created for each user that tried to login in the defined past time window and the edge between two nodes is a feature that is shared by two different accounts with two different account owners. A fraud ring/cluster is created once multiple nodes are identified to be accessed from the same actor who is not the account owner of any account present in the cluster, thus the same label is propagated throughout the cluster. But, for each cluster the identified fraudulent nodes are very few and are not directly linked to all nodes accessed by the same fraudulent actor. Therefore, we utilize the associated attributes with each node to find similar nodes accessed by the same actor.
\newline
\newline
Once a node is identified as fraudulent or non-fraudulent with high confidence (based on prior information) the node is assigned a label. For every node with a fraud label, the label is also propagated to the direct links as every node connected to the fraudulent node sharing the same attribute has high probability of being fraudulent as well. Thus, forming a set of ${m}$ clusters with labels ${L = \{l^j, j = 0, 1, 2, .., m\}}$ assigned to each cluster. The non-fraudulent cluster is assigned label 0 and the remaining labels represent group of accounts that are targeted by a specific actor forming a cluster for each actor that tried breaching the system. These labeled clusters form a small portion of the dataset and majority of the nodes remain unclassified. We provide these labeled nodes with their features as input to the model which tries to learn similarities between the labelled cluster nodes and the unlabelled nodes and propagates each cluster adding the most similar nodes to the cluster. All nodes belonging to the same cluster are either fraud attempts with the same behavior or from the same source.

\section{Network Architecture}

In this framework, we use Spectral Graph Convolution Network for node clustering. Spectral GCNs make use of the Eigen-decomposition of graph Laplacian matrix to implement information propagation. The Eigen-decomposition helps us understand the graph structure, hence, classifying graph nodes. Kipf \& Welling \cite{kipf2017semisupervised} define goal of GCNs to learn a function of node features on a graph which takes as input: 
\begin{itemize}
\item A feature description for each node, summarized in a ${N \times D}$ feature matrix ${X}$ where ${N}$ is the total number of nodes and ${D}$ is the number of input features.
\item An adjacency matrix ${A}$ representative of the description of graph structure in matrix form.
\end{itemize}
and produces ${Z}$, a ${N \times F}$ feature matrix as a node-level output.
\newline
\newline
Our linked fraud detection GCN model is comprised of 3 Graph Convolution layers. ReLU activation is applied to each Graph Convolution layer. Inputs to the network include adjacency matrix and node features and the final layer takes features returned from the previous hidden layer and outputs ${k}$ probabilities for each fraud cluster. Each graph convolution layer in the model can be represented as a non-linear function
\begin{equation}
  H^{(l+1)}=f(H^{(l)},A)
\end{equation}
Equation 1 defines a simplistic overview of the layer, but multiplication with the Adjacency Matrix ${A}$ would only incorporate feature vectors of the neighboring nodes but not the node itself. Thus, a self loop needs to be added to incorporate node feature vector. We also normalize Adjacency Matrix ${A}$ as multiplication with A may increase the scale of feature vector by large margins. To solve this Kipf \& Welling \cite{kipf2017semisupervised} proposed to following modifications to the propagation rule:
\begin{equation}
  f(H^{(l)},A) = \sigma(\hat{D}^{-\frac{1}{2}}\hat{A}\hat{D}^{-\frac{1}{2}}H^{(l)}W^{(l)})
\end{equation}
with ${\hat{A}=A+I}$, where ${I}$ is the identity matrix and ${\hat{D}}$ is the diagonal node degree of of ${\hat{A}}$
\newline
\newline
We utilized Deep Graph Library (DGL) \cite{wang2020deep} to define our Graph Convolution Network . DGL distills the computational patterns of graph neural networkss into few generalized sparse tensor operations suitable for extensive parallelization. By advocating graph as the central programming abstraction, DGL can perform optimizations transparently. We created a 3-layer GCN (as shown in Figure \ref{network}) with randomly initialized weights. We then input the adjaceny matrix and node features into our model. GCN performs three propagation steps during forward pass and convolves all nodes up to 3 hops away, i.e. 3rd order neighborhood of every node producing an embedding of these nodes that represent community-structure of the graph. The output layer then provides ${k}$ normalized embeddings resembling the probabilities of a node belonging to each community and the node is considered to belong to the community with highest probability.

\section{Experiment}
Most Account Takeover attacks can be identified by links between multiple activities. These links can either be strong or direct links like sharing the same attribute or weak like with similar behavior pattern. Solutions like community detection algorithms have been effective in identifying fraud via strong links whereas supervised classification algorithm have worked well in just binary classification of fraud learning a generic behavioral pattern. In this section, we provide a comparison between GCNs that utilize both strong and weak links to identify fraudulent communities with both Louvain label propagation algorithm \cite{7161493} for community detection and XGBoost algorithm for binary fraud classification.

\subsection{Dataset}
We experiment with two datasets to demonstrate our model performance for both binary classification and community detection. For binary classification, we randomly generate 34 nodes representing 34 accounts with 78 edges representing links between accounts. We create synthetic feature vectors with 5 attributes for each node representing login behavior like shared cookie count, mouse movement velocity, typing speed, shared ip count and login velocity. Data is generated such that login behavior for 11 accounts that were taken over is slightly different from the feature vectors representing non-fraud activities.
\newline
\newline
Community detection problem uses a larger dataset with 2356 nodes and 3560 edges, with 50 attributes representing behavioral pattern for fraudulent activities like attribute sharing counts, login, IP address, typing and movement velocities, userAgent data, geographical region data, email domain risk scores, email patterns etc. This dataset contains 4 different communities with 3 fraudulent communities with nodes of similar fraudulent behavior in one community and 1 community with non-fraudulent nodes. The communities with fraudulent nodes are much smaller than the non-fraudulent community representing real-world imbalance in fraud dataset where the fraudulent nodes are much less than the non-fraudulent nodes, ${\sim5\%}$  fraud-rate in this dataset. We test GCN, XGBoost and Cluster Label Propagation on both datasets and compare performance.

\subsection{Binary Fraud Classification}
With binary fraud classification, GCN identifies communities with fewer labeled  community nodes as given training data points (1 node per class for Binary Classification dataset). GCN during its training phase learns embeddings for each node which can be easily clustered together to identify communities. We also run Louvain Label Propagation (LLP) community detection algorithm supplying it just the graph to identify strongly linked graph nodes to detect communities. For fair comparison, we train an XGBoost model with the same number of nodes and node features as the GCN model (only 2 nodes per community) to predict fraud probability for the remaining nodes but just like other traditional classification algorithms XBBoost requires large training data and is not able to converge on a small dataset with only 34 nodes.


\begin{table}[t]
\caption{Binary Fraud Classification Results}
\label{binary_table}
\vskip 0.15in
\begin{center}
\begin{small}
\begin{sc}
\begin{tabular}{lcccr}
\toprule
Model & Recall & Precision & Time (seconds)\\
\midrule
GCN & 90.90\%  & 76.92\% & 0.011\\
LLP & 81.81\% & 69.23\% & 0.081\\
XGBoost & - & - & - \\
\bottomrule
\end{tabular}
\end{sc}
\end{small}
\end{center}
\vskip -0.1in
\end{table}

\begin{figure}[ht]
    \begin{center}
    	\fbox{\includegraphics[scale=0.11]{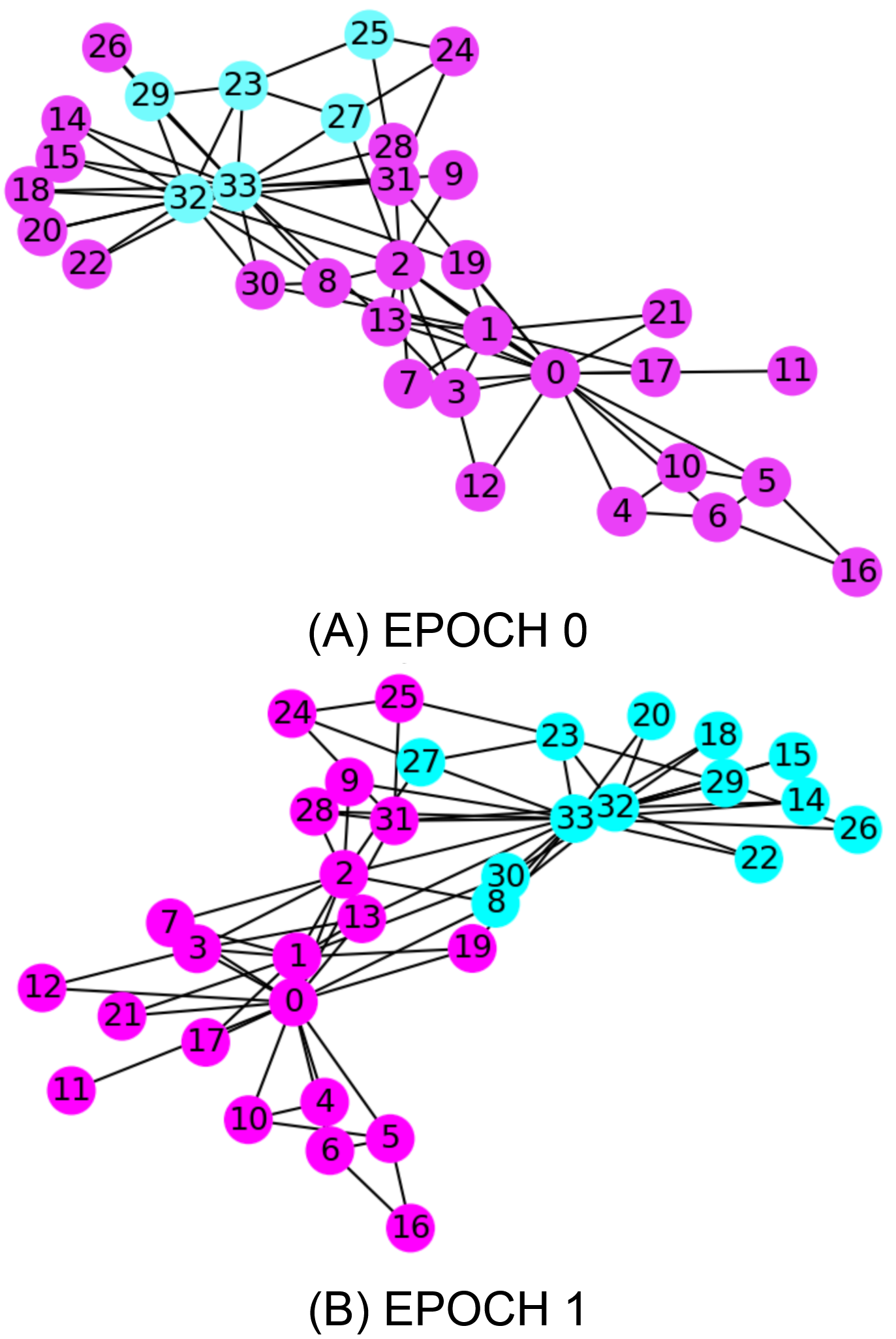}}
    	\caption{Louvain Label Propagation, pink nodes represent non fraudulent nodes and the blue nodes represent fraudulent nodes}
    	\label{llpepochs}
    \end{center}
\end{figure}

\begin{figure*}[ht]
\vskip 0.2in
  \centering
  \fbox{ \includegraphics[width=12cm]{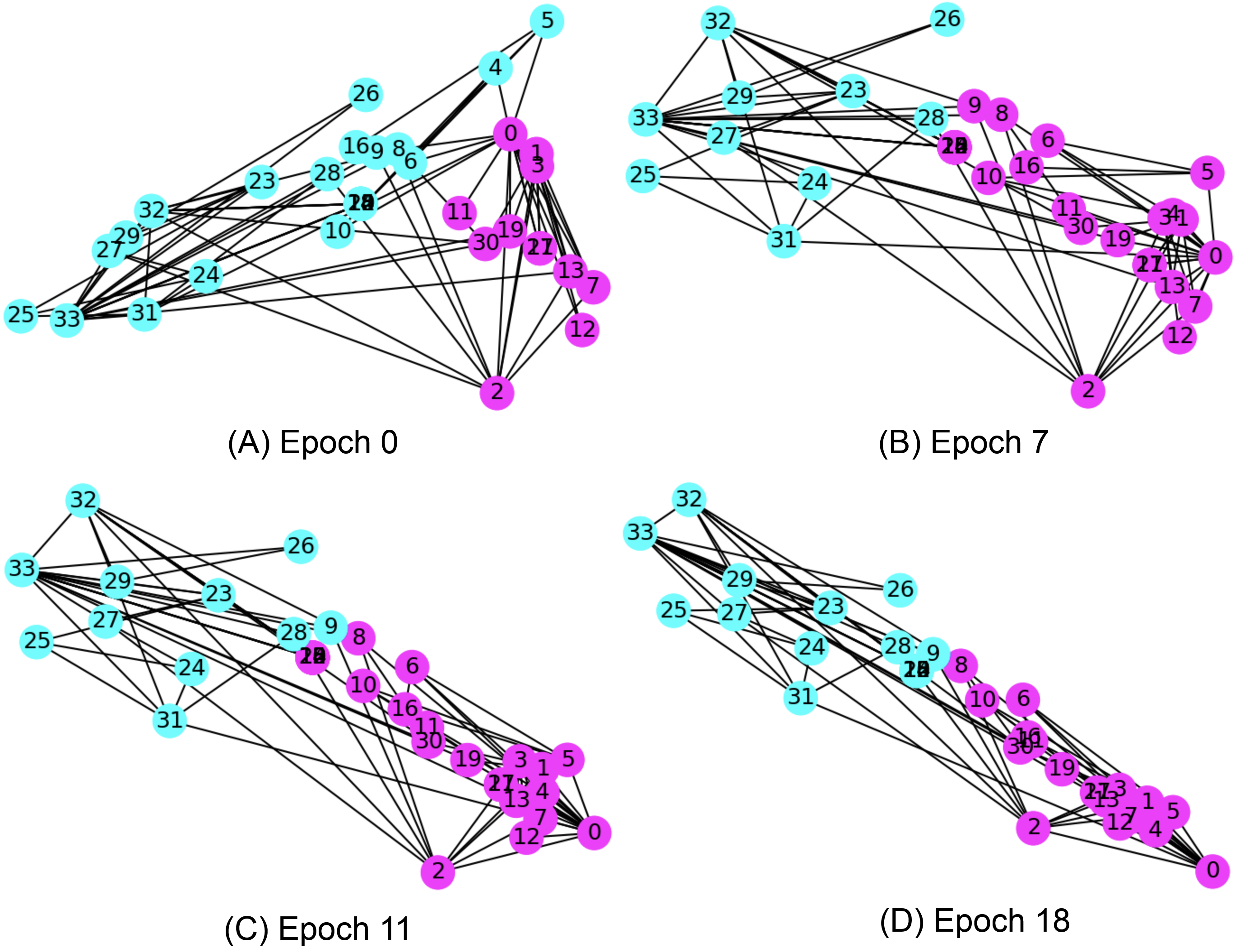}
  \rule[-.5cm]{0cm}{0cm} \rule[-.5cm]{0cm}{0cm}} 
  \caption{Graph Convolution Model Propagation, pink nodes represent non fraudulent nodes and the blue nodes represent fraudulent node}
  \label{gcnepochs}
  \vskip -0.2in
\end{figure*}

As shown in Table \ref{binary_table}, we calculate the precision and recall values for each model which informs about the amount of fraud identified by the models and also provide an idea of false positives and false negatives in the predictions. GCN were able to identify fraudulent nodes with higher precision and recall compared to community detection solution. XGBoost on the other hand, require much larger training dataset to learn fraud behavior. Even though GCN performs better in terms of solution quality it takes longer to converge and identify node clusters. GCN is able to achieve optimal performance after 20 epochs (Figure \ref{gcnepochs}) whereas Louvain Label Propagation algorithm saturates after 2 propagation epochs (Figure \ref{llpepochs}). In the next subsection, we try each model on much larger data to identify if GCNs can still achieve the same performance with faster convergence compared to LLP.

\subsection{Multi Fraud Community Detection}
The dataset for multiple community detection is much larger compared to binary classification in the previous section. We create a randomly generated graph with 2356 nodes and 3560 edges. Each node is associated with a feature vector of 50 attributes generated using PyTorch Embedding \cite{NEURIPS2019_9015} function for some nodes such that these nodes are divided into 4 different communities with nodes in the same community having similar behavior. Each fraudulent community represents either fraud from the same source or actor, or fraud of similar kind. We use SMOTE \cite{Chawla_2002} to create more feature vectors to increase the count of nodes to resemble same community and add noise to the vectors to make classification fair. We then manually assign these feature vectors to nodes in order to obtain 4 distinct communities, 3 fraudulent communities with 219, 196 and, 141 nodes respectively and 1 community  with 1800 nodes which represents the non-fraudulent community to mimic real-world distribution of fraud.
\newline
\newline
We provide 50 nodes from each community to GCN model to learn community behavior and identify the nodes closest embeddings. We train a multi-class XGBoost model with same training dataset and use the trained model to predict and assign a community to each node. For fair comparison we only train XGBoost model with 200 examples with 50 nodes from each community which were used to train the GCN model as well. We also use the same graph to train the Louvain Label Propagation community detection model to identify communities in the graph but as the LLP model is unsupervised and is not aware of total communities present in the dataset, it may identify more communities than the base knowledge.

\begin{table}[t]
\caption{Binary Fraud Classification Results}
\label{multi_table}
\vskip 0.15in
\begin{center}
\begin{small}
\begin{sc}
\begin{tabular}{lcccr}
\toprule
Model & Communities & Recall & Precision &  Time\\
\midrule
GCN & 4 & 92.61\%  & 84.40\%  & 0.718\\
LLP & 38 & - & - & 1.337\\
XGB & 4 & 72.41\% & 62.02\% & 0.371\\
\bottomrule
\end{tabular}
\end{sc}
\end{small}
\end{center}
\vskip -0.1in
\end{table}

As shown in Table \ref{multi_table}, we are able to achieve highest Precision and Recall with GCN model by a large margin. Although XGBoost model is faster, but due to fewer samples in training dataset XGBoost is not able to achieve optimal performance in terms of detection accuracy. The Louvain Propagation Model on the other hand is not able to identify the 4 communities at all but instead identifies much smaller 38 communities which makes it difficult to calculate performance metrics. LLP model is also much slower as the data size increases and needed 9 propagation epochs to saturate compared to 30 epochs taken by GCN to fit in about 0.619 seconds.

\section{Conclusion}
We conclude that for fraud problems that can be solved by detecting linked fraud either by sharing attributes or by identifying similar behavior, Graph Convolution Networks provide an ideal solution. GCNs require fewer samples to learn fraudulent behavior and identify activities with similar behavior. As we see in the experiment section, traditional Supervised Machine Learning Classification models require large data to train and learn fraud patterns in order to perform well which in case of fraud is not always available in such abundance. GCNs require very few examples of known fraud and can easily identify associated activities in comparable solution time. The semi-supervised approach of GCN provide them an upperhand compared to the unsupervised approach like Community Detection solution when it comes to identifying linked fraud as some prior knowledge of fraud is already available which can be leveraged to find all similar fraudulent activities. Also, compared to the clustering algorithms whose solution time complexities exponentially grow with the increase in dimension and quantity of data, GCNs require much less time to converge even with the increase in underlying data.

\section{Future Work}
In this paper, we demonstrated linked fraud identification for Account Takeovers as the linked fraud can be easily depicted as a Graph based problem and is more intuitive to understand and solve. Our proposed framework can be applied to a wide variety of problems in the fraud domain, such as, Anti-Money Laundering, Transaction Fraud, Check or ACH kiting, Identity thefts etc. Many such problems can be represented as a Graph. For example, money movement in case of Money Laundering and kiting is usually visualized as a Directed Acyclic Graph and problems like Identity Theft and Transaction fraud often have multiple traces of shared attributes or similar behavior, also known fraud in all such cases are usually low in volume therefore making them an ideal problem to be solved using Graph Convolution Networks.

\bibliography{example_paper}
\bibliographystyle{icml2021}

\end{document}